\documentclass[prl,twocolumn,superscriptaddress,amsfonts,amssymb,showpacs,a4]{revtex4}

\usepackage{graphicx}%
\usepackage{dcolumn}
\usepackage{amsmath}
\usepackage{dcolumn}
\usepackage{hyperref}
\usepackage{times}
\usepackage{xcolor}
\usepackage{soul}
\usepackage{bm}

\usepackage{draftcopy}
\draftcopySetGrey{0.7}

\makeatletter
\def\btt#1{\texttt{\@backslashchar#1}}%
\DeclareRobustCommand\bblash{\btt{\@backslashchar}}%
\makeatother

\newcommand{\G}{$\bar{\Gamma}$}
\newcommand{\X}{$\bar{X}$}

\newcommand{\dz}{$d_{3z^2-r^2}$}
\newcommand{\dxz}{$d_{xz/yz}$}

\newcommand{\FePF}{CeFeAs$_{1-x}$P$_x$O$_{0.95}$F$_{0.05}$}
\newcommand{\FePx}{CeFeAs$_{1-x}$P$_x$O}
\newcommand{\FeP}{CeFeAs$_{0.7}$P$_{0.3}$O}

\newcommand{\Sr}{SrFe$_2$P$_2$}

\usepackage{color}

\begin{document}
\sethlcolor{yellow}

\title{How chemical pressure affects the fundamental properties of rare-earth pnictides: an ARPES view}

\author{M. G. Holder}
\affiliation{Institut f\"ur Festk\"orperphysik, Technische Universit\"at Dresden, D-01062 Dresden, Germany}

\author{A. Jesche}
\affiliation{Max-Planck-Institut f\"ur Chemische Physik fester Stoffe, D-01187 Dresden, Germany}

\author{P. Lombardo}
\author{R. Hayn}
\affiliation{Institut Mat\'eriaux, Micro\'electronique et Nanosciences de Provence, Facult\'e St. J\'er\^ome, 13397 Marseille, France}

\author{D. V. Vyalikh}
\email{vyalikh@physik.phy.tu-dresden.de}
\affiliation{Institut f\"ur Festk\"orperphysik, Technische Universit\"at Dresden, D-01062 Dresden, Germany}

\author{K. Kummer}
\affiliation{European Synchrotron Radiation Facility, 6 Rue Jules Horowitz, B.P. 220, F-38043 Grenoble Cedex, France}

\author{S. Danzenb\"acher}
\affiliation{Institut f\"ur Festk\"orperphysik, Technische Universit\"at Dresden, D-01062 Dresden, Germany}

\author{C. Krellner}
\affiliation{Max-Planck-Institut f\"ur Chemische Physik fester Stoffe, D-01187 Dresden, Germany}

\author{C. Geibel}
\affiliation{Max-Planck-Institut f\"ur Chemische Physik fester Stoffe, D-01187 Dresden, Germany}

\author{E. D. L. Rienks}
\affiliation{Helmholtz-Zentrum Berlin f\"ur Materialien und
Energie GmbH, Elektronenspeicherring BESSY II, D-12489 Berlin,
Germany}

\author{S. L. Molodtsov}
\affiliation{European XFEL GmbH, Albert-Einstein-Ring 19, D-22671 Hamburg, Germany}

\author{C. Laubschat}
\affiliation{Institut f\"ur Festk\"orperphysik, Technische Universit\"at Dresden, D-01062 Dresden, Germany}

\date{\today}

\begin{abstract}

Angle-resolved photoelectron spectroscopy, supplemented by theoretical calculations has been applied to study the electronic structure of heavy-fermion material CeFePO, a homologue to the Fe-based high-temperature superconductors, and \FeP~, where the applied chemical pressure results in a ferromagnetic order of the 4$f$ moments. A comparative analysis reveals characteristic differences in the Fe-derived band structure for these materials, implying a rather different hybridization of valence electrons to the localized 4$f$ orbitals. In particular, our results suggest that the ferromagnetism of Ce moments in \FeP ~is mediated mainly by Fe 3\dxz orbitals, while the Kondo screening in CeFePO is instead due to a strong interaction of Fe 3\dz orbitals.

\end{abstract}


\maketitle

Rare-earth pnictides of the form $RE$Fe$Pn$O ($RE$: Rare Earth, $Pn$: As or P) have attracted considerable interest in the last few years due to the competition between magnetic and superconducting (SC) properties. Both are closely related to the unique topology of their Fermi surfaces. Hole and electron-like sheets around the $\Gamma$ and $M$ points, respectively, enable the adjustment of nesting conditions by charge carrier doping \cite{Mazin09,Kamihara08a,Zhao08}. At the same time, chemical pressure induced through the substitution of P by As allows to manipulate the band structure near the Fermi level ($E_F$)\cite{Vildosola08}.

In Ce-based iron pnictides, 4$f$ related excitations emerge directly at $E_F$, independent of any doping, and strongly influence the phenomenology. But while the 4$f$ interaction competes with SC in the heavy-fermion (HF) system CeFePO \cite{Bruning09,Holder10} there seems to be no remarkable influence of $4f$ hybridization on the ground state properties in the isoelectronic compound CeFeAsO \cite{Zhao08}. Several recent publications aim to figure out the correlation between SC, magnetism and $f-d$ hybridization by a systematic variation of the latter. In Ce-based systems external pressure is supposed to increase $f-d$ hybridization due to the decreasing lattice constant. In CeFePO this leads to a stabilization of the HF ground state \cite{Zocco11}, while in CeFeAsO$_{0.9}$ the SC at ambient pressure gets suppressed at a pressure of $\sim$ 5 GPa, likely because the Ce valence exceeds some critical value \cite{Yamaoka10}. 
Similar to applying external pressure, chemical pressure induced by P-As substitution can be used to modify the lattice constants in \FePx ~\cite{Luo10}.
This leads to suppression of the HF state and the onset of ferromagnetic order of the 4$f$ moments at $x$=0.95 \cite{Luo10}. For $x<0.3$ Ce-order switches from ferromagnetic to antiferromagnetic, along with the onset of a spin-density wave (SDW) ordering of the Fe spins. For the analogue EuFe$_2$As$_{2-x}$P$_x$ series a similar phase diagram can be drawn \cite{Jeevan11} but SC is found in coexistence with antiferromagnetic ordering of the $RE$ moments. In Ce based compounds this kind of behavior was reported for the doped series CeFeAs$_{1-x}$P$_x$O$_{0.95}$F$_{0.05}$  \cite{Luo11}, but recently also in CeFeAs$_{1-x}$P$_x$O single crystals ~\cite{Jesche}.

For all these series a reduced $f-d$ hybridization is expected due to the increasing lattice constants with increasing arsenide content. But at the same time local density approximation (LDA) studies of LaFeP$_{1-x}$As$_x$O  \cite{Vildosola08} show that the band structure close to $E_F$ is also strongly modified: Bands that are derived from the Fe 3\dz ~orbitals were found to be shifted to higher binding energy. On the other hand it had been shown in Ref.~\cite{Holder10}, that in particular these bands strongly interact with 4$f$ states close to $E_F$. However, this interaction is supposed to be effective, only as long as these states remain close to $E_F$. In fact HF behavior is suppressed, when $E_F$ is lifted by F doping \cite{Luo11}.

It is the aim of the present study to investigate the relation between details of the band structure and the strength of the $f-d$ hybridization in oxypnictides. To this end the results of angle-resolved photoelectron spectroscopy (ARPES) on the heavy-fermion compound CeFePO, previously reported in Ref.~\cite{Holder10}, are compared to the data of \FeP , where the applied chemical pressure results in a ferromagnetic order of the 4$f$ moments, but maintaining the system close to an antiferromagnetic phase transition. The lattice constants of \FeP ~are increased relative to CeFePO by 2.4~$\%$ perpendicular to and by 1.5~$\%$ within the Fe plane and a reduced spatial overlap of the $f$ and $d$ orbitals and thus reduced hybridization is expected. It is shown, however, that the energy shift of Fe 3\dz ~related bands has a much stronger influence on the phenomenology of these compounds. Based on the present results a mechanism is proposed that traces back the transition from Kondo screening to 4$f$ ferromagnetism to the competing interaction of the Ce 4$f$ orbitals with Fe 3\dz ~and 3\dxz ~related bands, respectively.

The photoemission experiments were performed at the $1^3$-ARPES setup at BESSY as described in Ref.~[\onlinecite{Inosov08}], at temperatures around 10\,K, on single crystals grown from Sn flux as specified in Ref.~[\onlinecite{Krellner08}]. The analysis of ARPES data throughout this paper leans on results previously published for CeFePO \cite{Holder10}. Horizontally and vertically polarized (HP, VP) light and photon energies of $h\nu$=112\,eV (Fano antiresonance) and 121\,eV (Fano resonance) are used to determine the symmetry of and $4f$ contributions to the electron bands, respectively.   


\begin{figure}
 \centering
 \includegraphics[width=80mm,clip=]{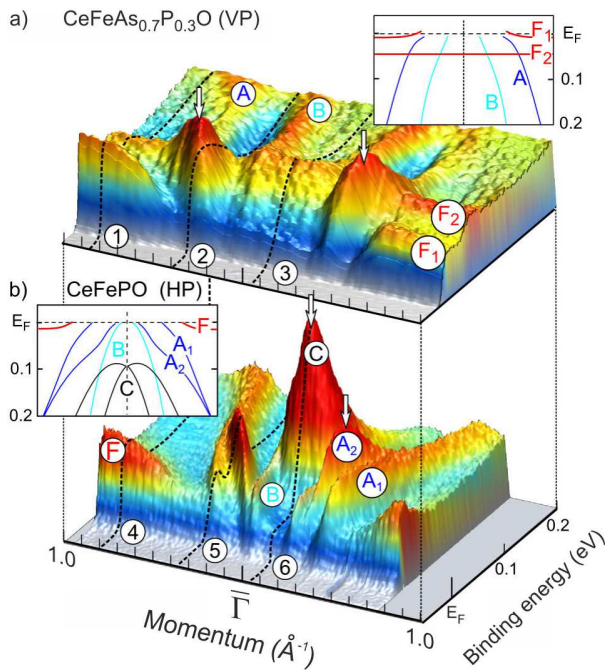}
 \caption{(Color online) Overview of ARPES results at h$\nu$=121\,eV along the \G -\X direction of a) \FeP ~at VP and b) CeFePO at HP \cite{Holder10}. Dashed lines indicate EDCs that are presented in Fig. 2. Insets in a) and b) show schematic drawings of the band structure, extracted from the experimental data.}
    \label{ARPES}
\end{figure}

To give an overview of the electronic structure and the Ce 4$f$ contribution to the valence band the most significant ARPES data of \FeP ~and CeFePO are presented in Fig.~\ref{ARPES} a) and b). Both parts of the figure show data that are obtained at $h\nu$=121\,eV, where 4$f$ emissions are enhanced due to the 4$d\rightarrow$4$f$ Fano resonance. Fig.~\ref{ARPES} only shows the measurement for that polarization of incident light, where the strongest $f$ contribution to the valence band of the respective compound were found: VP for \FeP ~and HP for CeFePO, respectively. 

The different dependence on polarization already points to a different symmetry character of valence band states that are involved in $f-d$ interaction and will be discussed later in more detail. In order to emphasize the rather weak $f$ contributions in the spectra of \FeP , the background signal had to be partly removed in this representation. A detailed discussion of the as-measured data at both polarizations is given later on for the energy distribution curves (EDCs) indicated by the dashed lines in Fig. 1. 

As illustrated in the inset in Fig.~\ref{ARPES} a) the valence band structure of \FeP ~close to $E_F$ comprises two bands, named $A$ and $B$. In addition two further $f$ related flat structures named $F_1$ and $F_2$ occur in the close vicinity of the Fermi level. The intensity of structure $F_1$ vanishes close to those values of $\vec{k}_{||}$ where band $A$ crosses the Fermi level, while the intensity of the latter increases strongly [white arrows in Fig.~\ref{ARPES} a)]. These maxima, that indicate a 4$f$ contribution to band $A$ due to the interaction with the 4$f$-derived structure $F_1$ are only observed at VP. This is in contrast to the observations made in Ref.~\cite{Holder10} on CeFePO. As displayed in Fig.~\ref{ARPES} b) the strongest $f$ contribution can be found in CeFePO to a band below the Fermi level, named $C$, that was ascribed in Ref.~\cite{Holder10} to be formed by mainly Fe 3\dz ~orbitals. At VP this emission vanishes nearly completely. Further $f$ contributions can be found to band $A_2$ that is, like band $A_1$, supposed to be mainly composed of the Fe 3\dxz ~orbitals. Similar to the ARPES data of \FeP ~a $f$-related structure $F$ occurs running closely parallel to the Fermi level. 

\begin{figure}
 \centering
 \includegraphics[width=80mm,clip=]{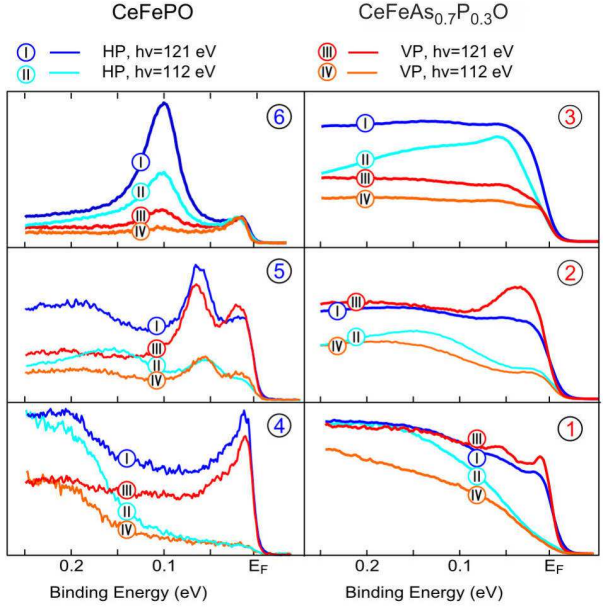}
\caption{(Color online) Comparison of the EDC along several cuts of the Brillouin zone. Arabic numbers refer to the dashed lines in Fig.~\ref{ARPES}. Top: EDC at \G , middle: Fermi level crossing of band $A$ and $A_1$ respectively and bottom: cut through the peak along the Fermi level.}
    \label{EDC}
\end{figure}

The different polarization dependence of the 4$f$ emission in both compounds points to a different symmetry character of the valence band states, that interact with the $f$ level. The labels of the bands in Fig.~\ref{ARPES} already anticipate an assignment of the bands observed in CeFePO and \FeP , that shall now be further justified by a detailed comparison of the as-measured data. Fig. \ref{EDC} compares the EDC at those $\vec{k}_{||}$ vectors that had been indicated by dashed lines in Fig.~\ref{ARPES}. In each of the displayed panels, the data are shown for HP [ \MakeUppercase{\romannumeral 1}) 121\,eV, \MakeUppercase{\romannumeral 2}) 112\,eV] as well as for VP [ \MakeUppercase{\romannumeral 3}) 121\,eV, \MakeUppercase{\romannumeral 4}) 112\,eV]. The 4$f$ emission for the specific polarization can be inferred from the difference between the EDC at the two photon energies.

The topmost panels in Fig.~\ref{EDC} show the EDCs exactly at \G , where actually the most relevant difference is found between the both compounds. A peak at 0.1\,eV binding energy (BE), corresponds to band $C$ in CeFePO. It gains intensity at resonant photon energy upon excitation with HP. From this behavior its origin from the Fe 3\dz ~orbitals, strongly hybridized with $f$ states has been concluded in Ref. \cite{Holder10}. In \FeP ~on the other hand no such strong peak is observed. At h$\nu$=121\,eV and VP a peak at 0.05\,eV and a shoulder at 0.15\,eV BE are found. The energy of the former one coincides with the BE of the structure $F_2$, but evidence for an interaction of this feature with the parabolic valence band $B$ as would be reflected e.g. by an increased intensity is not observed. Switching to HP both peaks are superimposed by an intense background signal.

Panels in the middle of Fig.~\ref{EDC} show the polarization dependence of bands $A_1$, $A_2$ and $A$. According to Ref.~\cite{Holder10} they are mainly derived from Fe 3\dxz ~orbitals. For CeFePO a cut at the Fermi level crossing of band $A_1$ has been chosen. Band $A_2$ is reflected by the peak at around 0.07\,eV BE. Emission from both electron bands are enhanced at resonance, but with opposite polarization dependence. Band $A_2$ is stronger at HP, pointing to some admixture of \dz ~orbitals. Mixing of \dxz ~with \dz ~orbitals is particularly expected for the crystal surface. Since the \dz ~orbitals are not observed close to the Fermi level in \FeP , a band similar to $A_2$ is not found there. A sizable $f$ contribution to bands $A_1$ of CeFePO and $A$ of \FeP , respectively, can only be observed at VP. This means that band $A_1$ in CeFePO and band A in \FeP ~are derived from the same orbitals. 

Panels in the bottom of Fig.~\ref{EDC} finally compare cuts through the $f$-derived Fermi level features of both compounds. In CeFePO a clearly visible peak develops at both polarizations. In \FeP ~instead two comparably weak structures on top of the background are observed. Indeed, the decreased $f-d$ hybridization in the As doped compound suggests much less intensity of the Fermi level peak in \FeP ~than in CeFePO (see e.g. Ref.~\cite{Hayn01}). Moreover the emissions nearly vanish in \FeP ~at HP but are slightly enhanced at VP. Thus, it roughly resembles the behavior of band $A$. The binding energy of the structure $F_2$ at around 0.06\,eV is close to the crystal field splitting of 4$f$ levels in CeFeAsO, as reported in Ref.~\cite{Chi08}. In CeFePO the Fermi level peak is strong at both polarizations and a shoulder around 0.04\,eV indicates also the existence of a satellite structure at, however, lower BE than observed in \FeP . 

From the results of the ARPES experiments as described above one can conclude that in both compounds the hybridization with localized 4$f$ orbitals relies on two kinds of valence band states, namely those derived from Fe 3\dz ~and 3\dxz ~orbitals, respectively. ARPES data of CeFePO indicate strong $f$ contribution to a valence band of Fe \dz ~character and comparably weak contribution to a band of Fe \dxz ~character. In \FeP ~only a contribution to the \dxz -derived band could be observed. This behavior can be interpreted on the basis of LDA calculations in Ref.~\cite{Vildosola08}: with increasing distance of pnictide atoms from the iron layer, bands of \dz character shift to higher BE. As a result, the interaction with the 4$f$ derived peak at $E_F$ is expected to become less effective as it is actually observed in \FeP .

In order to gain theoretical insight into the delicate interplay between 4$f$ and 3$d$ states in CeFePO, a simple approach has been proposed in Ref.~\cite{Holder10}, based on the periodic Anderson model (PAM). There, it was solved by dynamical mean field theory (DMFT) with the non-crossing approximation (NCA) as impurity solver. This model comprised only one single valence band that is to be hybridized with the 4$f$ state, and indeed was able to reproduce the characteristic behavior of band $C$ and the Fermi level peak $F$ in Fig.~\ref{ARPES} b). In the following we show that the same model is also suitable to describe the behavior of band $A$ in \FeP ~and band $A_1$ in CeFePO, too. To this end the width of the valence band was increased to $W$=1.8\,eV, its center shifted by 0.1\,eV to lower BE ($\epsilon_d$=0.6\,eV) and the $f-d$ hybridization strength set to $t_{df}$=0.25\,eV, while Coulomb repulsion $U_{ff}$=7\,eV, and the energy of the bare $f$ level $\epsilon_f$=2\,eV BE remained constant.

\begin{figure}
 \centering
 \includegraphics[width=80mm,clip=]{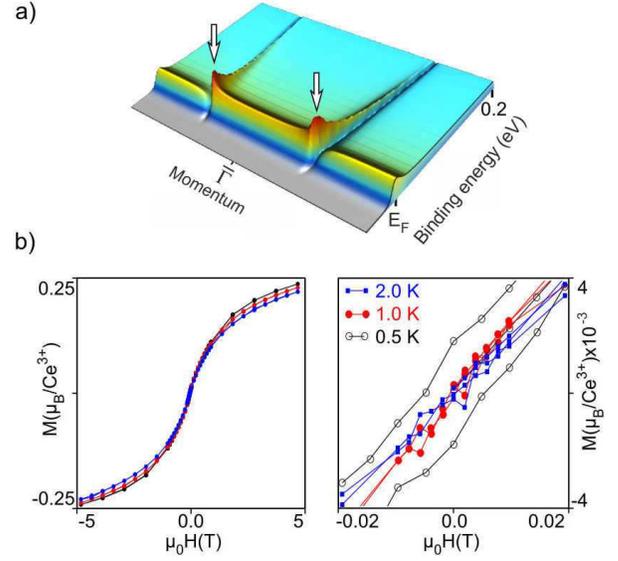}
\caption{(Color online) a) $f$ spectral function of a model band structure roughly reflecting the behavior of band $A$, solved with DMFT method. b) magnetization of CeFePO as a function of the applied field perpendicular to the iron plane.}
\label{BandStr}
\end{figure}

The resulting 4$f$ spectral function is shown in Fig.~\ref{BandStr}. It comprises one non-dispersing peak along the Fermi level that develops due to $f-d$ hybridization and the $f$ contribution to the valence band. (A further peak, so-called ionization peak, that occurs close to the binding energy of the bare $f$ level $\epsilon_f$ is not shown in this plot). At the crossing with the valence band the Fermi level peak vanishes, while band $A$ gains strong 4$f$ admixture. As observed in the ARPES data, the maximum $f$ contribution occurs at $\vec{k}_F$ [white arrows in Fig.~\ref{BandStr} a)], while intensity drops toward the \G ~point. It is interesting to note, that the hight of this maxima remains by one order of magnitude smaller than in the scenario for CeFePO, reported in Ref.~\cite{Holder10}. In contrast to the theoretical prediction, the experimentally observed intensity of feature $F_1$ vanishes around the \G ~point, while that of $F_2$ remains almost constant. Since crystal field splittings are not included in the model calculation, one should regard the theoretically obtained 4$f$ peak as a superposition of both features $F_1$ and $F_2$.

Altogether this means, that despite its simplicity, the proposed approach allows for a qualitative description of the different behavior of the \dz and \dxz bands in CeFePO. Even though quantitatively exact results might not be expected, the NCA seems to allow for further improvements in existing LDA plus DMFT methods, which so far had not been able to describe the experimentally observed Fermi level peak \cite{Pourovskii2008}. Note, that in the limit $U_{ff}\rightarrow\infty$ the non-dispersive part of the Fermi level peak disappears and only the $f$-contribution at the $E_F$ crossings, $\vec{k}_F$, remain visible reproducing results obtained with a simple approach to PAM for the same limit \cite{Danz05}. In the following we proceed with a qualitative discussion of the relation between band structure and magnetic behavior in \FePx .

A widely used description of magnetic coupling between localized moments utilizes the Ruderman-Kittel-Kasuya-Yosida (RKKY) mechanism \cite{Yosida1991}. Within this model the magnetic coupling between localized moments is mediated by valence band states, hybridized with the 4$f$ orbitals. Since such a hybridization is found in \FeP ~for band $A_1$, RKKY mediated ferromagnetism appears possible. However, a similar interaction for band $A$ in CeFePO gives also reason for ferromagnetic coupling. Indeed at least ferromagnetic fluctuations had been stated in polycrystalline samples of CeFePO \cite{Bruning09}. In order to find hints for a ferromagnetic interaction in the single crystalline samples the magnetization as a function of the applied field has been measured in the present study for several temperatures well below the Kondo temperature of about 10\,K. The magnetization as a function of the field perpendicular to the Fe plane is shown in Fig.~\ref{BandStr} b). At temperatures of 1.0 and 2.0\,K no hysteresis occurs within experimental accuracy. But at 0.5\,K a spontaneous magnetic moment in the range of $10^{-3}\mu_B$ is found. Recent muon-spin relaxation experiments on CeFePO down to T=20 mK revealed static, short-range, bulk magnetic order of the Ce-ions, in strong support of our findings. A more detailed discussion of the spin relaxation results will be reported elsewhere \cite{last}.

Thus, one can anticipate, that in both compounds the same magnetic interaction, even though with different strength, should exist. It is mediated presumably by \dxz ~orbitals, which form band $A$ in \FeP ~and band $A_1$ in CeFePO, respectively. However, in the latter case the strength of the magnetic interaction strongly reduced, due to the competition between RKKY mediated magnetism and the Kondo-screening of the local moments, that is usually  treated in terms of the dependence of characteristic temperatures on hybridization strength \cite{doniach}. The exponential dependence of the Kondo temperature favors the paramagnetic heavy-fermion ground state as hybridization increases. Since the lattice constant is reduced in CeFePO, increased overlap and thus stronger hybridization is expected for \dxz ~orbitals. However, the energy shift of the \dz ~band towards to $E_F$, that incorporates a further strong interaction is supposed to be much more important. Thus, the hybridization of these orbitals might be responsible for the formation of heavy fermions in CeFePO, while the interaction of \dxz ~orbitals (band $A_1$) gives rise to the ferromagnetic interaction, although with much reduced strength. 

If the As content is further increased the ferromagnetic order of the 4$f$ moments changes to an antiferromagnetic one. Thus one could argue an intricate relation to SDW ordering of Fe spins in \FePx , which is accompanied by the formation of an energy gap, when electronic states belonging to different sheets of the Fermi surface are linked by nesting conditions \cite{Overhauser}. Such a SDW transition was mapped by ARPES for \Sr ~in Ref. \cite{Hsieh09} which shows that it is accompanied by energy lowering of bands around the \G ~point. If band $A$ in \FePx ~was also involved in such an interaction (for $x<$ 0.3) the proposed ferromagnetic RKKY mechanism could be changed to an antiferromagnetic one. However, it was shown that for $x$=0, internal fields of the  Fe-magnetism competes with the RKKY interaction between the 4$f$ moments ~\cite{Hayn01, Maeter, A1}, leading probably to a much more complex interplay between 4$f$ and 3$d$ physics. 

Notwithstanding, in \FePF ~as well as in EuFe$_2$As$_{2-x}$P$_x$ the transition from ferromagnetic to antiferromagnetic order of rare earth spins is accompanied by onsetting SC instead of SDW. The latter suggests, that at least in these compounds the magnetic transition is triggered by further changes of the band structure due to chemical pressure as it is also stated in Ref.~\cite{Jeevan11}. This could mean, that SC can only exist if hybridization of \dxz ~orbitals as it was discussed throughout this paper is suppressed, stating the relevance of these orbitals for superconducting interaction. For a detailed discussion one needs to take into account the whole Fermi surface and not only a single cut as it is done in the present case. In particular sheets around the $M$ point might also contribute to magnetic coupling. But this is so far not resolved experimentally.   

In summary a mechanism for the magnetic transitions in \FePx , based on the ARPES and model calculation results of CeFePO and \FeP ~is proposed. The bands formed by the Fe 3\dz ~orbitals reveal strong hybridization with $f$ states giving rise to heavy-fermion behavior. The shift of these orbitals to higher BE as chemical pressure is applied, circumvent this interaction in favor of much weaker hybridization of the \dxz ~orbitals. The latter is argued to be responsible for ferromagnetic interaction involving the RKKY mechanism. 

This work was supported by DFG project VY 64/1-1, GE602/2-1 and LA655/12-1. The
authors would like to thank S. Borisenko group for support at "1$^3$-ARPES" beam line at BESSY, Ch. Klausnitzer for help with magnetization
measurements and K. Matho for fruitful discussions.

\end{document}